\documentclass[aps,eqsecnum,preprint,floats,epsf,epsfig,nofootinbib]{revtex4}
\usepackage{graphicx}
\usepackage{epsfig}

\begin{document}
\rightline{BNL-HET-04/20}   
\def\be{\begin{eqnarray}}
\def\en{\end{eqnarray}}
\def\non{\nonumber}
\def\la{\langle}
\def\ra{\rangle}
\def\pp{{\prime\prime}}
\def\nc{N_c^{\rm eff}}
\def\vp{\varepsilon}
\def\hep{\hat{\varepsilon}}
\def\a{{\cal A}}
\def\B{{\cal B}}
\def\c{{\cal C}}
\def\d{{\cal D}}
\def\e{{\cal E}}
\def\p{{\cal P}}
\def\t{{\cal T}}
\def\B{{\cal B}}
\def\P{{\cal P}}
\def\T{{\cal T}}
\def\C{{\cal C}}
\def\A{{\cal A}}
\def\E{{\cal E}}
\def\CP{$CP$~}
\def\up{\uparrow}
\def\dw{\downarrow}
\def\vma{{_{V-A}}}
\def\vpa{{_{V+A}}}
\def\smp{{_{S-P}}}
\def\spp{{_{S+P}}}
\def\3bar{{\bf \bar 3}}
\def\6bar{{\bf \bar 6}}
\def\10bar{{\bf \ov{10}}}
\def\ov{\overline}
\def\Lqcd{{\Lambda_{\rm QCD}}}
\def\lsim{ {\
\lower-1.2pt\vbox{\hbox{\rlap{$<$}\lower5pt\vbox{\hbox{$\sim$}
}}}\ } }
\def\gsim{ {\
\lower-1.2pt\vbox{\hbox{\rlap{$>$}\lower5pt\vbox{\hbox{$\sim$}
}}}\ } }
\def\jpsi{J/\psi~}
\def\BAR{\overline}
\def\xba{\overline}
\def\fm{{\cal M}}
\def\fl{{\cal L}}
\def\ufs{\Upsilon(5S)}
\def\ufour{\Upsilon(4S)}
\def\xcp{X_{CP}}
\def\ynotcp{Y}
\vspace*{-.5in}
\def\etap{\eta^\prime}
\def\rhobar{\overline\rho}
\def\etabar{\overline\eta}
\def\g5{\gamma_5}
\def\bea{\begin{eqnarray*}}
\def\eea{\end{eqnarray*}}
\def\bo{B^0}
\def\bs{B_s}
\def\acps{a_{CP}^{B \to X_s \gamma}}
\def\acpd{a_{CP}^{B \to X_d \gamma}}
\def\PR{Phys. Rev. D}
\def\PRL{Phys. Rev. Lett}
\def\PL{Phys. Lett. B}
\def\NP{Nucl. Phys. }
\def\ZP{Zeit. fur. Physik }
\def\RMP{Revs. of Mod. Phys. }
\def\IJMP{Int. Jour. of Mod. Phys. }
\def\np{\newpage}
\def\bib{\bibitem}
\def\noi{\noindent}
\def\uglu{\hskip 0pt plus 1fil minus 1fil}
\def\uglux{\hskip 0pt plus .75fil minus .75fil}
\def\slashed#1{\setbox200=\hbox{$ #1 $}
    \hbox{\box200 \hskip -\wd200 \hbox to \wd200 {\uglu $/$ \uglux}}}

\def\slpar{\slashed\partial}
\def\sla{\slashed a}
\def\slb{\slashed b}
\def\slc{\slashed c}
\def\sld{\slashed d}
\def\sle{\slashed e}
\def\slf{\slashed f}
\def\slg{\slashed g}
\def\slh{\slashed h}
\def\sli{\slashed i}
\def\slj{\slashed j}
\def\slk{\slashed k}
\def\sll{\slashed l}
\def\slm{\slashed m}
\def\sln{\slashed n}
\def\slo{\slashed o}
\def\slp{\slashed p}
\def\slq{\slashed q}
\def\slr{\slashed r}
\def\sls{\slashed s}
\def\slt{\slashed t}
\def\slu{\slashed u}
\def\slv{\slashed v}
\def\slw{\slashed w}
\def\slx{\slashed x}
\def\sly{\slashed y}
\def\slz{\slashed z}

\centerline{\large\bf Search for New Physics at a Super-B Factory$^\circ$}  
\bigskip
\bigskip
\centerline{\bf Thomas E. Browder$^1$ and 
Amarjit Soni$^2$}
\medskip
\centerline{$^1$ Department of Physics, University of Hawaii, Honolulu,
Hawaii 96822} 
\medskip

\medskip
\centerline{$^2$ High Energy Theory Group, Brookhaven National
Laboratory} \centerline{Upton, New York 11973}
\footnotetext{Invited plenary talk at the Workshop on High Energy
Physics Phenomenology (WHEPP8), Indian Institute of Technology,
Mumbai (India), Jan. 5-16, 2004} 
\medskip

\bigskip
\bigskip
\centerline{\bf Abstract}
\bigskip
The importance of a Super-B Factory in the search for New Physics,
in particular, due to CP-od phase(s) from physics beyond the Standard Model 
is surveyed. The first point to emphasize is that we know now
how to directly measure all three angles of the unitarity triangle
very cleanly, {\it i.e. without theoretical assumptions} with
{\it irreducible theory error $ \lsim 1\%$};
however this requires much more luminosity than is currently available
at B-factories. Direct searches via penguin-dominated hadronic
modes as well as radiative, pair-leptonic and semi-leptonic
decays are also discussed. {\bf Null tests of the SM}
are stressed as these will play a crucial role especially if
the effects of BSM phase(s) on B-physics are small.

\eject

\section{Introduction and Motivation}
The asymmetric B-factories at KEK and SLAC have performed remarkably
well.
The accurate measurement of CP asymmetry
in $B^0 \to \jpsi K^0$ was significant for
a variety of reasons. For one thing, it constitutes the first evidence
of CP violation outside of the K-system. Unlike the case of K-decays,
though, in $B \to \jpsi K^0$ the asymmetry
is large\cite{belle_psik,babar_psik}, i.e. O(1), which is
about three orders of magnitude larger than what is found in K-decays.
Quantitatively, the measured asymmetry provides a striking
confirmation of the CKM-paradigm\cite{ckm}
as it is in very good agreement with
indirect determinations based on the SM. As such these measurements at the
asymmetric B-factories represent an important milestone in our
understanding of CP violation phenomena.

Indeed, these studies also show that the CKM phase is the dominant
contributor to the observed CP asymmetry and the effect of any
beyond the Standard Model (BSM) CP-odd phase(s)
(we will collectively denote these as $\chi_{BSM}$), even if they
exist, in $B \to \jpsi K^0$ must be small\cite{as_paschos}. This great
success of the B factories also now entails a concern for the future
as it implies  that the effect of $\chi_{BSM}$ in B-physics
may be small and their detection may be experimentally quite
challenging.  At the same time it is important to reemphasize that
there are very good reasons to suggest that
$\chi_{BSM}$ must exist.

In extensions of the SM, as a rule, it is difficult to
avoid new phase(s). Given that three families of quarks exist,
in the context of the SM, a CP-odd phase in the CKM matrix
occurs naturally. In fact, although it is not impossible
to arrange extensions of the SM in such a way that the CP-odd
phase of the CKM matrix is zero, these tend to be contrived
and not natural.

From the perspective of modern quantum field theory there is nothing
sacred about CP asymmetry. As more particles (fermions, gauge bosons
or scalars) are introduced in extended models new CP-odd
phases arise\cite{baes_pr}. More explicitly, this can be seen in specific
extensions such as two Higgs doublets\cite{tdl,sw}, LRS\cite
{pati,kiers_lr}, SUSY \cite{gnr} or models with warped
extra dimensions\cite{aps}. In
the case of two Higgs doublet models (2HDM) with natural
flavor conservation there are three neutral Higgs whose exchanges
entail a new CP-odd phase\cite{baes_pr,bernreuther}.
In general, in minimal LRS models
based on the gauge group $SU(2)_L \times SU(2)_R \times U(1)$ 
there can be as many
as six new phases. Similarly SUSY can have tens of
new phases\cite{gnr}.
Thus while the SM - CKM phase is completely natural, at the same time,
there is no good reason to think that more CP -odd phases do not exist.

Furthermore, repeated investigations have suggested that the CKM
phase is unable to account for baryogenesis. So far this has provided
the only clue for the need of BSM source(s) of CP violation. Indeed,
at the same time,
it seems that extended models such as 2HDM, LR symmetry, SUSY or
Warped Extra Dimensions
may be able to account for this crucial requirement of any
model of CP violation.

While there are very strong reasons to think that BSM CP phase(s)
exist, there is no reliable guidance as to the size of the effects
they cause in B-physics. Indeed, the asymmetry they cause may be small
even if the underlying CP-odd phases are not small. In this context,
the SM itself teaches a valuable lesson. We know now that the CKM phase
is O(1) yet the asymmetry it produces in K decays is vanishingly small.
The indirect and direct CP violation parameters in $K_L$ decays are
$\epsilon_K \approx 10^{-3}$ and $\epsilon'_K \approx 10^{-6}$.
In fact, in decays of the top quark the CKM phase is expected
to cause even smaller asymmetries\cite{ehs,baes_pr}
(than in K-decays) and there is virtually no hope
of ever being able to detect them in laboratory experiments.
It is also well known
that in charm decays the SM causes extremely small CP violating
asymmetries. Thus, it is only in B-decays that the CKM
phase causes large asymmetries. It therefore stands to reason that
the CP asymmetries, due to BSM phase(s), in B-decays, need not be large
and may well be quite small.

Interestingly, there are some indications that the SM description
of time-dependent CP asymmetry in penguin dominated $b \to s$
decays, such as $B \to \phi (\eta', \pi^0...) K^0$,
may not be adequate. Indeed TDCP asymmetries O(10-20\%) due
to BSM sources cannot at this 
point be ruled out~\cite{mg_ichep04,ys_ichep04}.  

On the other hand, if the CP asymmetries in B-physics 
caused by $\chi_{BSM}$ are as small
as they were in K-decays i.e. $O(10^{-3})$ then their detection will
undoubtedly require very large fluxes of B-mesons in a clean
environment.
Assuming a Br of $O(10^{-3})$, since it is difficult to find modes
(that may be useful for this purpose)
with larger Br, it is easy to see that the detection of asymmetries
of $O(10^{-3})$ requires $\ge 10^{10}$ B-mesons i.e. a super-B
factory~\cite{as_izu} as well as precise control of experimental systematics. 

The situation with respect to
$\chi_{BSM}$ is somewhat reminiscent of $\nu$ mass and
oscillations. Their was never any good reason to think that
$m_\nu = 0$ just as there are no good reasons to think that
$\chi_{BSM} =0$. For decades the only experimental indication of the
possible need for $m_\nu \neq 0$ was the deficit of solar
$\nu$'s. Similarly, the fact that it 
is difficult to account for baryogenesis 
with the CKM-phase serves as a beacon for the search
for $\chi_{BSM}$. The search for $\nu$ mass and $\nu$ oscillations
took decades. In fact, the $\Delta {m^2}$ region had to be persistently
lowered by about three orders of magnitude just in the
past two deades before neutrino oscillations
and $m_\nu \neq 0$ were established !
We can only hope that nature would be kinder
for $\chi_{BSM}$ in B-physics but we cannot count on that.

What we can count on is improved determination of the angles of the UT.
The important point is that we know now methods which allow us to
extract
all the angles of the UT (and not just $\beta$)
\cite{notation}
very cleanly i.e. with intrinsic theory errors
that are very small, {\it i.e.} $\le 1\%$,
but that require substantially more B mesons than
are available at a B-factory.

If the BSM source(s) cause only small deviations from the SM
in B-Physics, then in addition to precision determination of the angles
of the UT, searches for NP via {\it null tests} of the SM could also
play an important role. Furthermore, progress in our calculational
prowess would also be highly desirable.

\section{Strategies for Improved Searches of New Physics
at a Super-B Factory}
In light of the important findings of the two
asymmetric B-factories, the strategies for searching
for new physics (NP) may be subdivided into three broad
categories as follows:\\
\begin{itemize}
\item Indirect Searches with theory input
\item Indirect Searches without theory input: Elements of a Pristine UT
\item Direct Searches (TDCP, DIRCP, PRA, TCA ..wherever applicable)
especially in arenas where the SM predicts vanishing asymmetries.

\noindent In particular, {\bf NULL TESTS} of the CKM-paradigm become
extremely important especially if $\chi_{BSM}$ leads to
small deviations in B-physics from the predictions of the SM.
\end{itemize}
\subsection{Indirect Searches with theory input}

In the Wolfenstein representation, the four parameters of the CKM
matrix are $\lambda, A,\rho$ and $\eta$. Of these,
$\lambda = 0.2200 \pm 0.0026$ \cite{PDB}, $A \approx 0.850 \pm 0.035$
are known quite precisely; $\rho$ and $\eta$ still need to be 
determined accurately. Efforts have been underway for many
years to determine these parameters. The angles $\alpha, \beta, \gamma$,
of the UT can be determined once one knows the 4-CKM parameters.

A well studied strategy for determining these from experimental data
requires knowledge of hadronic matrix elements.  
Efforts to calculate several
of the relevant matrix elements on the lattice, with increasing
accuracy, have been underway for past many years. 
A central role is played by the following four 
inputs~\cite{ans_ckm,marti_ckm,fitter_ckm}: 

\begin{itemize}
\item $B_K$ from the lattice with $\epsilon_K$ from experiment
\item $f_B\sqrt{B_B}$ from the lattice with $\Delta m_d$ from
experiment
\item $ \xi $ from the lattice with $\frac{\Delta m_s}{\Delta m_d}$
from experiment
\item $\frac{b \to u l \nu}{b \to c l \nu}$
from phenomenology, HQS, lattice $+$ experiment
\end{itemize}

As is well known, for the past few years, these inputs lead to the important
constraint: $ \sin 2\beta_{SM} \approx 0.70 \pm 0.10$
which was found to be 
in very good agreement with direct experimental determination,
(by B-factories + CDF +...)
via CP asymmetry measurements from $b\to c \bar{c} s$ decays (such
as $B \to \jpsi K^0$)
$sin 2 \beta_{expt} =0.726 \pm 0.037$~\cite{mg_ichep04,ys_ichep04}\

Despite severe limitations (e.g. the so-called quenched approximation)
these lattice inputs provide
valuable help so that with B-Factory measurements one arrives at the
very important
conclusion that in $B \to \jpsi K^0$ the CKM-phase is
the dominant contributor; any NP contribution 
is unlikely to be greater than about 15\%.  

What sort of progress can we expect from the lattice in the next 
several years in these (indirect) determination
of the UT? To answer this it is of some use to
see the pace of progress of the past several years.
Table~\ref{lat_tab} shows how lattice calculations
of matrix elements around 1995~\cite{lat95_as}
yielded (amongst other things) $\sin 2\beta \approx 0.59 \pm
0.20$, whereas the corresponding error decreased to around 
${\pm 0.10}$ around 2001~\cite{ans_ckm,marti_ckm,fitter_ckm}. 
In addition to $\beta$,
such calculations also now constrain $\gamma (\approx 60^\circ)$
with an error of around $10^\circ$.

\begin{table}
\begin{center}
\begin{tabular}{|c|c|c|}
\hline
Quantity & Old Fit(Lat'95) & New Fit(BCP4 '01) \\
\hline
$V_{ub}/V_{cb}$ & $0.08 \pm .02$ & $0.085 \pm .017$ \\
\hline
$V_{cb}$ & $0.04 \pm .005$ & $0.04 \pm .002$ \\
\hline
$f_B \sqrt{\hat B_B}$(MeV) & $237 \pm 65$ & $230 \pm 50$ \\
\hline
$ \xi $ & $1.16 \pm .10 $ & $1.20 \pm .10$  \\
\hline
$\hat B_K$ & $0.85 \pm .21$ & $0.86 \pm .15$ \\
\hline
$\sin 2\beta_{SM}$ & $0.59 \pm .20$ &$ 0.70 \pm .10$     \\
\hline
$\eta$ & $0.32 \pm .10 $ & $0.30 \pm .05$      \\
\hline
$V_{td}/V_{ts}$ & $0.22 \pm .05$ &$ 0.185 \pm .015$ \\
\hline
\end{tabular}
\end{center}
\caption{Illustration of progress from lattice calculations towards
constraining CKM-parameters} 
\label{lat_tab}
\end{table}

\vskip 0.2in
There are three important developments that should help lattice
calculations in the near future:\\
\begin{enumerate}
\item{} Exact chiral symmetry can be maintained on the lattice.
This is especially important for light quark physics.   
\item{} Relatively inexpensive methods for simulations
with dynamical quarks (esp. using improved staggered fermions) 
have become
available. This should help overcome limitations of the 
quenched approximation. 
\item{} About an order of magnitude increase in computing power
is imminent. Another factor of about 2-3 is quite likely in the 
forthcoming
years.
\end{enumerate}

In 5 years or so
errors on lattice determination of CKM parameters should
decrease 
appreciably, perhaps by a factor of 3.
So the error in $\sin 2\beta_{SM} \pm 0.10 \to
\pm 0.03$;
$\gamma \pm 10^\circ \to 4^\circ$ etc.
While this increase in accuracy is very welcome,
there are good reasons to believe, experiment will
move ahead of theory in direct determinations
of unitarity angles in 3-5 years. (At present, experiment is already ahead
of theory for $\sin 2 \beta$).

\subsection{Indirect searches without theory input:
Elements of a Superclean Unitarity Triangle}

The starting point is to recognize that measurement
of the angle $\beta$ by the B-factories with essentially
no theoretical assumptions has ushered in a new area.

The basic idea here is very simple. One should use methods that
are extremely `` clean" {\it i.e.} require no
theoretical assumptions and directly measure all the angles
of the unitarity triangle. Thereby through redundant
measurements of this type one can 
test CKM-unitarity and look for inclusive signals of $\chi_{BSM}$.

In spirit, this is a generalization of the great success of
the B-factories in directly measuring the angle
$\beta$ with time-dependent CP asymmetry studies
in $B^0 / \bar B^0 \to \jpsi K^0$.  At the moment the
error in the measurement of $\beta$ is around
5\%. However, this is expected to improve quite
rapidly as more data is accumulated. The important point
for this discussion is that the intrinsic theory error
for the method being used is $\le 1\%$.

Let us briefly recall that\\
\begin{itemize}

\item{} Direct CP studies of $B^{\pm} \to ``K^{\pm}" + D^0, \bar D^0$
give $\gamma$~\cite{gw,ads,ggsz}.  

\item{} Time dependent CP studies in $B^0 \to ``K^0" D^0, \bar D^0$
gives $\gamma$ {\it OR } $\alpha$ {\it AND} $\beta$~\cite{gl,ans}.
\end{itemize}

Although the $\beta$ determination from $B^0 \to D^0 K^0$ is
not competitive~\cite{ans} with the $B \to J/\psi K^0$ method, 
it is still useful as it provides
a good check of the CKM-paradigm. Note, in particular,
that for $B^{\pm}$ and for $B^0$ these methods
for extracting angles of the UT
using decays to $D^0$ final states are very clean
as they require no theoretical assumptions such
as isospin~\cite{ans_pristine}. 

Furthermore, time dependent and direct CP studies in all three
final states of $B \to \pi \pi$, $\rho \pi$, $\rho \rho$
should give a very good determination of $\alpha$~\cite{hq1}.

With these methods of direct determination
of the unitarity angles, a very important
criteria to bear in mind is the {\it irreducible theory error} (ITE).
In other words this is the intrinsic error coming 
from theoretical assumptions that 
these methods entail and even with
very large data samples it will be very difficult to
reduce this error.  The $ B \to K D$ methods for $\gamma$ are likely
to have the smallest ITE, perhaps O(0.1\%). The ITE for
$\beta$ with the $B \to \jpsi K^0$ mode is also expected to
be less than a percent. The $\alpha$ determination
is less clean due to EWP contamination, resonant substructure,
resonance-continuum
separation difficulties etc. However, when all three FS
($\pi \pi, \rho \pi,\rho \rho$)
are studied (given enough luminosity) it is quite
plausible that the remaining ITE for $\alpha$ also will be quite small,
{\it i.e.} O(1\%).

It is extremely important that we make use of the opportunity
afforded to us by as many of 
these very clean redundant measurements as possible.
In order to exploit these methods to their
fullest potential and get the angles with
errors of order ITE will require a Super-B Factory (SBF)~\cite{as_izu}.

The crucial point that cannot be overemphasized is that just
as the feasibility of a clean measurement of
$\beta$ was the central motivating factor for the
construction of the asymmetric B-factories
about a decade ago, we should understand that we now know
of methods that will allow us to cleanly and directly measure
the other two angles, $\gamma$ and $\alpha$.  
Although this motivation for SBF may not appear ``sexy"
to some, it is important to understand that it is based
entirely on facts; no theoretical assumptions, prejudices or
speculations are involved.
Therefore, this ought to be a very important
driving force for construction of a super B-factory as these methods
do require much larger luminosities than what the 
current B-factories can deliver.

Precision measurements of the three
angles in itself constitutes a strong enough reason for
a SBF, as it represents a great opportunity to precisely
nail down the fundamental parameters of the CKM paradigm.

\section{Direct Searches}

We focus mainly on the following types of direct searches:
\begin{enumerate}

\item{} Mixing-induced CP violation in radiative B-decays.
We begin the discussion with this interesting method for
searching for NP as it is relatively new and seems
very promising.

\item{} Mixing-induced CP violation in penguin 
dominated hadronic final states.
This has become very topical in the past two years
and is likely to remain an
important test of the CKM-paradigm for a long time. 
Indeed, CPV in $b\to s$ decays
has the distinction of providing a plausible
indication of a non-standard phase that
could be causing sizeable deviations in B-physics
from the expectations of the SM. 

\item{} Radiative B-decays, branching ratios
and direct CP. Although the Br measurement has played
an important role for almost a decade in limiting the parameter space
of NP, it is likely to become less effective.
However, direct CP searches are extremely important
to pursue for quite some time to come and 
will be accompanied, in any case, by more precise measurements of Br's.

\item{} Decays with leptonic pairs {\it e.g.}
$B \to X l^+ l^-$. Experiments are just beginning to make
sensitive measurements of the rate.
More accurate determination of the rate as well
as forward-backward asymmetries,
direct CP, including triple correlation asymmetries 
are clearly important.

\item{} Importance of tree dominated hadronic final states
that are especially sensitive to
a CP-odd phase from the charged Higgs sector.  

\item{} Semi-leptonic decays into final states with a $\tau$
lepton. These are especially suited to constraining 
the charged Higgs-sector. Furthermore, the importance
of the transverse polarization asymmetry as a powerful null test
of the CKM-paradigm is emphasized.  

\end{enumerate}

\subsection{Mixing Induced CP in Radiative B-decays}

While the use of the rate and the direct CP asymmetry measurements
in radiative decays of the B have received much attention
for a very long time\cite{hurth_rev}, discussions on
mixing-induced (time dependent) CP
in these modes are more recent~\cite{ags97}.   
This class of CP provides a very clean test of the SM
and is very sensitive to presence of
right handed currents of BSM origin\cite{ags97}.

The key point is that in the SM, the $\gamma$ in b decays is
predominantly LH
whereas the $\gamma$ in $\bar b$ decays is predominantly
RH. Mixing-induced CP cannot occur unless the $B^0$ and $\bar B^0$
decay to the same final state to enable them to interfere.
Thus in the SM TDCP asymmetry in radiative
B-decays are expected to be either very small ($b \to s$) or
completely negligible ($b \to d$).      

In the SM TDCP
in $B \to \gamma [\rho,\omega,K^*,..] \propto m_d/m_b$ or
$m_s/m_
b$ (in addition to including other suppression factors as
well).
BSM physics [e.g. LRSM, SUSY] can produce much larger asymmetries
as in those models the occurrence of  RH currents
does not necessarily suffer from the $m_d(m_s)/m_b$ suppression
factor. Implications for these reactions in BSM scenarios
have been studied recently in many papers\cite{hou0110}.

In general, (for $q = s,d$)\\
\begin{equation}
H_{eff} = -\sqrt8G_F\frac{em_b}{16\pi^2}F_{\mu \nu}[\frac{1}{2}F_L^q
\bar q \sigma^{\mu \nu} (1 + \g5) b + \frac{1}{2}F_R^q \bar q
\sigma^{\mu \nu} (1 - \g5) b]
\end{equation}

In the SM, $\frac{F_R^q}{F_L^q} \approx \frac{m_q}{m_b}$
In contrast, for example in a LR model,  
$\frac{F_R^q}{F_L^q}$ can be appreciably larger as the
presence of RH currents 
has a $m_t/m_b$
enhancement for $\frac{F_R^q}{F_L^q}$\cite{ags97}.

\subsubsection{Time Dependent CP Asymmetry in $B(t) \to M^0 \gamma$}

For a state tagged as a B rather than a $\bar B$ at $t=0$
and with $CP|M^0>=\xi|M^0>$; with $\xi=\pm 1$ :\\
\begin{eqnarray}
A(\bar B\to M^0\gamma_L)&=&A\cos\psi e^{i\phi_L}~,\nonumber\\
A(\bar B\to M^0\gamma_R)&=&A\sin\psi e^{i\phi_R}~,\nonumber\\
A(B\to M^0\gamma_R)&=&\xi A\cos\psi e^{-i\phi_L}~,\nonumber\\
A(B\to M^0\gamma_L)&=&\xi A\sin\psi e^{-i\phi_R}~.
\end{eqnarray}

Here $tan \psi=\frac{F_R^q}{F_L^q}$ and $\phi_{L,R}$ are
CP-odd weak phases. Thus, with $\phi_M$ as the mixing phase,
$\Gamma(t)\equiv\Gamma(B(t)\to M^0\gamma)$,
\begin{equation} 
\Gamma(t)=e^{-\Gamma t}|A|^2[1+\xi
\sin(2\psi)\sin(\phi_M-\phi_L-\phi_R)\sin(\Delta mt)]~.
\end{equation}  
This leads to a time-dependent CP asymmetry,
\begin{equation}
A(t)\equiv {\Gamma(t)-\bar\Gamma(t)\over \Gamma(t)+\bar\Gamma(t)}
=\xi\sin(2\psi)\sin(\phi_M-\phi_L-\phi_R)\sin(\Delta mt)~.
\end{equation}

In the SM:\\
\begin{eqnarray}
{\rm for}~\bo&:&~~~\phi_M=2\beta ~, \nonumber\\
{\rm for}~\bs&:&~~~\phi_M=0~,
\end{eqnarray}
and
\begin{eqnarray}
{\rm for}~b\to s\gamma : ~~~\sin(2\psi)\approx {2m_s\over
m_b}~,~~~\phi_L=\phi_R
\approx 0~,
\nonumber\\
{\rm for}~b\to d\gamma : ~~~\sin(2\psi)\approx {2m_d\over
m_b}~,~~~\phi_L=\phi_R
\approx \beta~,
\end{eqnarray}
Thus, in the SM, :\\
\begin{eqnarray}
\bo \to K^{*0}\gamma &:& A(t) \approx (2m_s/m_b)\sin(2\beta)\sin(\Delta
mt)~,
\nonumber\\
\bo \to \rho^0\gamma &:& A(t) \approx 0~,
\nonumber\\
\bs \to \phi\gamma &:& A(t) \approx 0~,
\nonumber\\
\bs \to K^{*0}\gamma &:& A(t) \approx -(2m_d/m_b)\sin(2\beta)\sin(\Delta
mt)~,
\label{examples}
\end{eqnarray}
where the $K^{*0}$ is observed through $K^{*0}\to K_S
\pi^0$.
Therefore, the SM predicts a maximum of a 
few percent ($\approx 3\%$) TDCP
asymmetry in the $B^0 \to K^* \gamma$ mode whereas the asymmetry
in the $B^0 \to \rho \gamma$ mode ought to be completely negligble.

As an illustrative contrast with the SM, let us next consider
a simple LRSM based on the EW gauge group,
$G = SU(2)_L \times SU(2)_R \times U(1)$. As is well known
the left and right handed doublets of quarks and leptons
occur in this model completely symmetrically, e.g.
\bea
\left( \begin{array}{cc}
u\\
d\\
\end{array} \right)_{L,R}, 
\left( \begin{array}{c}
\nu_e\\
e
\end{array} \right)_{L,R}
\eea
This class of models
has many attractive features, e.g. the $\nu$ mass arises naturally.
Using the $K_L -K_S$ mass difference one obtains a rather stringent
bound $m_R \ge 1.5$ TeV\cite{bbs_lr}.
Given that $m_\nu \neq 0$ (and TeV is no longer
such an imposing scale as it seemed in the early 80's)
the model ought to be reconsidered as
an effective low energy theory\cite{kiers_lr}.
Taking,
$<\Phi>=
\left( \begin{array}{cc}
\kappa & 0 \\
0 & \kappa'
\end{array} \right)$
and setting $|\kappa'/\kappa|=m_b/m_t$ leads to the striking
simplification\cite{kiers_lr}:\\
$\Rightarrow$ The CKM angle hierarchy arises quite readily\\
$\Rightarrow (CKM)_R=(CKM)_L$\\
$\Rightarrow \delta_R=\delta_L$\\
endowing the model with a ``natural" origin for
so-called ``manifest" LR symmetry and considerable predictive power.

\begin{figure}[h]
\begin{center}
\hskip -1in
\epsfxsize=2.5in
\epsfbox{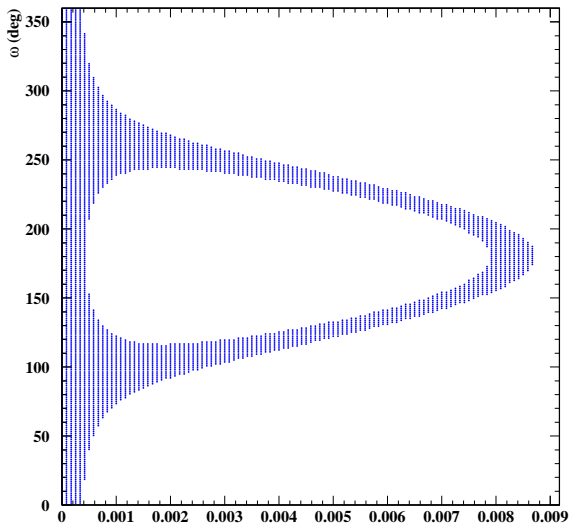}
\caption{
Presently allowed values of $\zeta$ and $\omega$ from
$Br (B \to X_s \gamma)$, deduced by setting EXP/SM $= 0.71 \pm 0.36$ (i.e.
to 90 \% CL), are included in the shaded area and in the blank internal
area. Only the shaded region would be allowed when agreement between
the SM prediction and experiment at the 10\% 
level is attained~\cite{ags_notes}. 
\label{ags_fig}}
\end{center}
\end{figure}

The $W_L-W_R$ mixing is described by\\
\begin{equation}  
\left( \begin{array}{c}
W_1^+\\
W_2^+
\end{array} \right)
=
\left( \begin{array}{cc}
\cos\zeta & e^{-i\omega}\sin\zeta \\
-\sin\zeta & e^{-i\omega}\cos\zeta
\end{array} \right)
\ \ \
\left( \begin{array}{c}
W_L^+\\
W_R^+
\end{array} \right)~.
\end{equation}  
Although $\zeta$ is small, $\le 3\times 10^{-3}$,
\cite{bs_moriond83,wolf84} that is
considerably offset by the helicity enhancement factor $m_t/m_b$.
Radiative B-decays previously examined in the LRSM showed,
\cite{fy94_12}

\bea
F_L\propto F(x)+ \eta_{QCD} + \zeta {m_t\over m_b} e^{i\omega}\tilde
F(x)
\eea

\bea
F_R\propto \zeta {m_t\over m_b} e^{-i\omega}\tilde F(x)~
\eea

%
%
%
\noindent where $x=(m_t/m_{W_1})^2,~\eta_{QCD}=-0.18$. Also
assuming $\displaystyle
\frac{BR(B \to X_s \gamma)_{EXP}}{BR(B \to X_s \gamma)_{SM}}
=1.0 \pm 0.1 \Rightarrow |\sin (2 \omega)| =0.67$
[see Fig.~\ref{ags_fig}]
one obtains the predictions for time-dependent CP shown in Table~\ref{ags_tab}

\begin{table}
\begin{center}
\hspace*{-.2in}
\begin{tabular}{|c|c|c|}
\hline
Process & SM & LRSM \\
\hline
$A(B \to K^* \gamma)$ & $2\frac{m_s}{m_b}
\sin 2\beta \sin(\Delta m_t)$ &
$\sin 2 \omega \cos 2 \beta \sin (\Delta m_t)$ \\
\hline
$A(B \to \rho \gamma)$ & $\approx 0$ & $\sin 2 \omega \sin (\Delta m_t)$
\\
\hline
\end{tabular}
\end{center}
\caption{Mixing-induced CP asymmetries in radiative exclusive
B-decays in the SM and in the LRSM.}
\label{ags_tab}
\end{table}

Thus, whereas in the SM negligible TD asymmetries are predicted, in the
LRSM
they can be O(50\%) even if $BR(B \to X_s \gamma)$ is in very good
agreement with the SM.

The rarer radiative decay $B \to \rho \gamma$ provides an even more
striking contrast between the predictions
of the SM and a model-with LR currents such as
the LRSM. In the SM, mixing-induced CP is even
more dramatically suppressed as the quark mass ratio
now gets replaced by $m_d/m_b$. In addition,
the CP-odd weak phase factor of $\sin 2\beta$ is replaced
by a factor of $O(\lambda^2)$ so that in
the SM, CP asymmetry is expected to be
$\le 10^{-3}$, making it a very useful (essentially) {\it null test}.
In the LRSM $B \to \rho \gamma$ can have a TDCP asymmetry of order
tens of percents.

\vskip 0.5in

Let us briefly mention the curent experimental effort
to search for this class of asymmetry in radiative B-decays.  
Both BABAR and BELLE have demonstrated the feasibility
of time dependent CP asymmetry measurement in $B \to K^* [K^* \to K_s
\pi^0] \gamma$~\cite{babar_ags,belle_ags}. 
With some 100-200 $\times 10^6$ B-pairs each,
they both obtain results consistent with zero with statistical
errors of $O(0.6)$~\cite{babar_ags,belle_ags,ys_ichep04}. 
In the next five years or so, 
it is expected that the luminosities will increase by 
factors of perhaps 5-10. This would reduce this error to perhaps
around $O(0.2)$. If a positive result of such a size
is seen then that would of course unambiguously imply
new physics. However, for the experiments to reach the sensitivity of the
predicted SM asymmetry, an important goal,
would require higher luminosity
only accessible to a SBF~\cite{as_izu,neubert_jan,belle_loi}.


\subsection{Search for $\chi_{BSM}$ via penguin dominated hadronic 
final states}

This test of the SM has received considerable attention
in the past year or so. Initially, both BELLE and BABAR
saw a large negative central value (with large errors) for
$\sin 2 \beta$ from pure penguin modes.
However, in Summer'03 with somewhat improved statistics
the BABAR central value shifted and became quite 
consistent with the SM while Belle's value became more precise
and less consistent with the SM. The  
present (Summer'04) experimental status 
is summarized in Table~\ref{penguin_tab}.
The combined result for $b \to s$ modes from Belle and BaBar
deviates by about $3.5 \sigma$ from the SM.

Recall that in the SM, $B^0$, $\bar B^0$ decays that are
dominated by $b\to s$ penguin transitions are expected to have
a negligible CP-odd phase\cite{gw97,ls97}.
Since the $B \to \jpsi K^0$-like ($b\to c \bar{c} s$) decays
also receive no CP-odd weak phase, the
time dependent CP asymmetry measurements of these two seemingly
significantly different FS should have the same CKM-phase
originating from $B^0-\bar B^0$ mixing, i.e. $\sin 2 \beta$.

Note though that the u-quark contribution inside
the penguin loop, or for that matter the corresponding tree
(b to u) contributions do, in principle, carry a non-zero
CP-odd weak phase, {\it i.e.} $\gamma$. To that
extent the time-dependent CP asymmetry measured via
b to s ``penguin dominated" modes may differ from that
seen in $B \to \jpsi K^0$ modes.
For many of the modes of interest, the 
tree contribution is color
and Cabibbo suppressed; for those cases one finds~\cite{ls97}
$T/P$ to be $O(\lambda^2) \approx 0.04$. Thus, theoretical
estimates of these deviations in the modes listed above seem to
be O(a few \%). Since light-quarks are involved
some dynamical enhancement may well make these deviations
somewhat larger. However, naively it seems difficult
for the $\sin 2 \beta$ measured via these ``penguin" modes
to deviate by more than 10\% from the value measured in
the $\jpsi K^0$ method.  

As was emphasized in \cite{ls97}, in addition to 
$B \to \phi K^0$ several other modes
in which the $T/P$ ratio is very small (i.e. less than a few percent),
such as $\eta'$ ($\pi^0$, $\omega$, $\rho^0$...) $K_S^0$ can all
be used for testing the CKM paradigm. In each of these modes the magnitude
of time dependent CP-asymmetry should equal
$\sin 2 \beta$ to a very good approximation. The sign of the expected
asymmetry in the SM can be fixed by the CP eigenvalue of
the final state. 

\subsubsection{Highlights of the current experimental status~\cite{tblp03,
mg_ichep04,ys_ichep04}}

Table~\ref{penguin_tab} summarizes the experimental studies 
of the time-dependent CP in penguin dominated modes.
At the moment, the evidence for a significant 
difference from
the $\jpsi K^0$ ({\it i.e.} the $b \to c \bar c s$) determination:
\noindent $\sin(2\phi_1)=0.726 \pm0.037$
(weighted
average)\cite{mg_ichep04,ys_ichep04} is not yet compelling.
However, each
experiment sees an interesting $\approx 2.5 \sigma$ effect,
when all such modes are combined.
For a cleaner theoretical interpretation, though, in the long run,
it is better to separate
modes that receive color-allowed tree (CAT) contributions
from those that do not receive any tree contribution
or at most receive a color-suppressed tree contribution.
(In Table~\ref{penguin_tab} the $K^+ K^- K_S^0$ mode may also be different
from the rest due to the fact that it is a three body mode.) 
The above results are based on a combined total
of about 500 $\times 10^6$ B-pairs between the two experiments. With 
the expected increase in luminosity by a factor of 5-10 in
the next few years, the current error (around 0.12) should
decrease to around 0.05 making it a very meaningful test.
Again, for a cleaner theoretical interpretation and
a decisive test the error on individual modes
should be reduced to below $O(\lambda^2 \approx 0.05)$. 
This is very likely
to require a SBF and should be one of the most interesting
applications of such a machine~\cite{as_izu,neubert_jan,belle_loi}.  

\begin{table}
\begin{center}
\hspace*{-.2in}
\begin{tabular}{|c|c|c|c}
\hline
Final State & Type of Tree & BELLE & BABAR \\
\hline
$\phi K^0$ & NT & $0.06 \pm 0.33 \pm 0.09$ &
$0.50 \pm 0.25^{+0.07}_{0.04}$\\
\hline
$\eta' K_s^0$ & CST & $0.65 \pm 0.18 \pm 0.04$ & $0.27 \pm 0.14 \pm 0.03$\\
\hline
$f_0K_s$ & CST& $-0.47 \pm 0.41 \pm 0.08$ & $0.95^{+0.23}_{0.32} \pm 0.10$\\ 
\hline
$\pi^0 K_s$ & CST & $0.30 \pm 0.59 \pm 0.11$ & $0.35^{+0.30}_{0.33}\pm0.04$\\
\hline
$\omega K_s$ & CST & $0.75 \pm 0.64^{+0.13}_{-0.16}$&  \\
\hline
$K^+ K ^-K_s^0$ & CAT & $0.49 \pm 0.18^{+0.17}_{-0.04}$ & $ 0.55 \pm 0.22 \pm
0.12$\\
\hline
Average & & $0.43^{+0.12}_{-0.11}$ & $0.42 \pm 0.10$\\ 
\hline
\end{tabular}
\end{center}
\caption{Experimental status of search of time-dependent
CP in some penguin-dominated
modes~\cite{mg_ichep04,ys_ichep04}.
NT means no tree, CST is color suppressed tree
and CAT is color allowed tree} 
\label{penguin_tab}
\end{table}

\subsubsection{Model Independent Remarks}

For a model independent discussion,
we can divide NP sources contributing to $B \to \phi
K_s$ into 2 types and discuss briefly the implications of each:\\

I. NP leads to modification of $b \to s$ form-factor(s)\cite{as97}:\\
\begin{equation}  
\Lambda^{bs}_\mu  =  \bar s_i T^a_{ij} [-iF(q^2)
(q^2\gamma_\mu-q_\mu\not q)L
+ m_b q_\mu \epsilon_\nu
\sigma^{\mu\nu} G(q^2) R] b_j
\end{equation}  

with
\bea
F(q^2) = e^{i\delta_{st}} F_{SM} +
e^{i\lambda_F} F_x;
\eea

\bea
G(q^2) =  G_{SM} + e^{i\lambda_G}G_x
\eea

\noindent where $\delta_{st}$ is the
strong phase generated
by the absorptive part resulting from the
$c\bar c$ cut for
$q^2>4m^2_c$\cite{bss79};
$\lambda_F$
and $\lambda_G$ are the CP-odd
non-standard phases.
For simplicity, the CKM phase in $b \to s$
is assumed to be negligibly small .
Of course,
$glu \to q \bar q$ interactions as
dictated by  QCD are always possible and are implied.
So, $glu \to s \bar s$ leads to
the $\phi K_S^0$ anomaly; but at the
same time has serious ramifications for $\eta' K_s$.
In fact, recall that such a BSM modification was
introduced to enhance the rate for $B \to \eta' X_s (K)$,
possibly leading to non-standard 
direct CP violation signals\cite{as97}.
Also note that, for example, $ gluon \to c \bar c , ...$ is
inevitable.

Thus, it is clear that this type of new physics
should lead to deviations from SM
in numerous channels, in particular, all FS with (net) $\Delta S=\pm 1$
are susceptible to effects of NP: rates, DIRCP, TDCP, TCA
should all be modified. The effects will not be restricted to
$\phi K^0$ but will also be present in $\phi K^{\pm}$,
$\phi K^*$ (TCA), $K \bar K K (X)$;$\pi^0 K_s$,
$\eta' K_s, \eta' K^{\pm}...;$
$\sin (2 \beta)$ via $D^+ D^-$ should NOT equal that
from $\jpsi K^0$; also DIRCP in $D_s D^-(D^0)$, TCA in $D_s^* D^*$...;
Similarly $\gamma X_s (K^*, K \pi...); l^+ l^- X_s (K, K^*, K
\pi...)$ should also show deviations at some level 
depending on the detailed implementation of the BSM.

II. Another model independent
way to incorporate NP is
to assume an effective 4-fermi interaction in the $b \to s \bar s s$
vertex:
\begin{equation} 
L_{4f}^{b3s} = G_{b3s} e^{i \chi_{b3s}} [\bar s \Gamma_{\mu} b]
[\bar s
\Gamma^{\prime}_{\mu} ].  
\end{equation}   
\noindent $G_{b3s}$ is the effective 4-fermi coupling, assumed real;
$\chi_{b3s}$ is the associated non-standard CP-odd phase.
This is much more restrictive and yet such NP
should effect not just TDCP in $B\to \phi K^0$ but also
DIRCP in $B\to \phi K^0 (K^{\pm}, K^*...)$ and TCA in $B\to \phi K^*$;
similarly $K \bar K K (X)$; $\eta' K_s (K^{\pm}, K^*)$
should receive BSM contributions.

Thus we can draw some general conclusions:
\begin{enumerate}
\item{} It is impossible to isolate NP only in TDCP in $B\to \phi K^0$
\item{}All channels affected by II are also affected by I
(but not vice versa)
\item{} Many of these NP effects will occur in the $B_s$
system as well; e.g. $\Delta m_s$,
TDCP in $\phi \phi$, $\phi \eta'(\eta)$ and TCA in
$\phi K \bar K (X)$. 

\end{enumerate}
\subsubsection{Some Implications of $BSM_s$ invoked to explain $\phi
K_s$}

Here are some 
illustrative examples that emphasize possible corroborative
evidence if one assumes that a large deviation from the SM is found in 
$b \to s$ TDCPV:  

I. Huang and Zhu\cite{huang_0307}
study
2HDM (Mod III) and find
TDCPA ($S_{\phi K}$) can occur with either sign but
DIRCPA $C_{\phi K}
> 0$\\

II. Raidal~\cite{mr_0208} in a LRSM
with a relatively low scale for $m_{W_R}$ and with at least
one new CP-odd phase found large TDCPA in $B \to K^* (\rho)
\gamma$;
CP violation in $B_s \to \phi \phi$ 
(also $\eta \rho$, $\pi^0
\rho$)\\

III.Hiller\cite{gh_0308}
and Atwood and Hiller\cite{ah_0307}
propose flavor-changing  $s Z' b$ with a complex coupling.
This leads to large non-standard effects in the Br 
and $A_{FB}$ of $b \to s l^+ l^-$; 
$B_s \to \mu \mu$; $\Delta m_s$\\

IV.Khalil and Kou \cite{kk_0307}
emphasized that SUSY can (interestingly) account for different asymmetries
in $B \to \phi K_S^0$ and $B \to \eta' K_S^0$. In particular,
they emphasize that the parity of the two final
states is not the same and as a result in BSM scenarios such
as SUSY $\phi K_S^0$ and $\eta' K_S^0$ can have different asymmetries.
In their SUSY scenario, DIRCP will occur even in $B^{\pm}$ decays;
non-standard helicity will arise in $b \to s \gamma$ and thus, for
example, TDCPA in $B \to K^* \gamma$ may also occur~\cite{romans_susy}.

\subsubsection{Summary on $B\to \phi K^0$}

\begin{itemize}
\item  Many beyond the SM scenarios
can accommodate fairly large deviations of the
asymmetry in $B\to \phi K^0$ from the SM expectation.
\item It is virtually impossible to confine 
the effects of a new phase to 
$B\to \phi K^0$; large TDCPA, DIRCP, TCA effects
should be seen in a multitude of channels. In particular,
TCA and other anomalous effects in $\phi K^*$,
$\pi^0 K_s$, $KKK (n \pi)$,
$\eta' K (n \pi)$, $\gamma K^* (n \pi)$, $l^+ l^- K(n \pi)$
should be vigorously studied.
\item Future experimental efforts should target definitive
measurements of asymmetry of $O(\approx theo. errors) \approx \lambda^2$
i.e. about 5\% in as many of these
individual channels as possible. Given a Br $\approx 10^{-5}$ and assuming 
a 10\% detection efficiency implies 
that about $10^{10} B \bar B$ pairs are required for a convincing
($5 \sigma$) signal i.e. a Super-B factory.
\item Modes that seem to be dominated by penguins
but that receive color-allowed tree contribution
{\it e.g.} $K^+ K^- K_s$ should not
be combined with those that only receive color-suppressed tree
contributions.
\end{itemize}

\subsection{Radiative B-Decays, Br and Direct CP}

Another very interesting rare B mode, whose importance has
been recognized for a very long time
\cite{hurth_rev} is $B \to X_s \gamma$.
Recall the current experimental status,
(World Ave.)
$Br (B \to X_s \gamma) = (3.34 \pm .38) \times 10^{-4}$
\cite{nakao_lp03}. In comparison, the
SM (NLO) predicts $(3.57 \pm 0.30) \times 10^{-4}$
\cite{bm_0207,hurth_rev}, which is in good agreement.

As is well known, this leads to important constraints on numerous
extensions of the SM such as,
2HDM's, supersymmetric or extra-dimension models,etc.\cite{hurth_rev}.
To further improve the theoretical prediction requires
NNLO calculations, a very demanding and challenging task.
Therefore, improvement in the experimental
determination of the Br may appear
somewhat unnecessary. However, larger data samples and improved
statistics are in any case essential for a better determination
of $a_{CP}^{B \to X_s \gamma}$, $Br(B \to X_d \gamma)$
and $a_{CP}^{B \to X_d \gamma}$, 
which are very well motivated.
The exclusive counterparts of these reactions 
also deserve continuing efforts.

In this context,
note that $a_{CP}^{B \to X_d \gamma}=-.004 \pm .051 \pm .038$
\cite{nakao_lp03}.
Thus, the current experimental limit on $\acps$ needs improvement by a factor
of 5-10 to reach sensitivity to the SM (i.e. $\acps \approx 0.6 \%$).
This should be possible at a SBF. Clearly precise measurements
of this asymmetry constitute a very important test of the SM.

Furthermore, let us  recall that
due to accidental cancellations, in 2HDMs $\acps$ is also $<0.6\%$,
however, it can be much larger in SUSY supergravity
inspired models\cite{tg_9812,ksw_0006}.

Interestingly, SM predicts $\acpd$ to be much larger ($\approx -16\%$)
\cite{ksw_0006}; see Table~\ref{ksw_tab}.

\subsubsection{Illustrative examples of constraints on models from
Br[$B \to X_s
\gamma$]}

For the past many years, Br($B \to X_s \gamma$) has been extremely
useful in constraining the parameter space of a wide variety of
non-standard models. Here are few examples that serve to illustrate
this point.

\begin{figure}[h]
\begin{center}
\hskip -1in
\epsfxsize=5in
\epsfbox{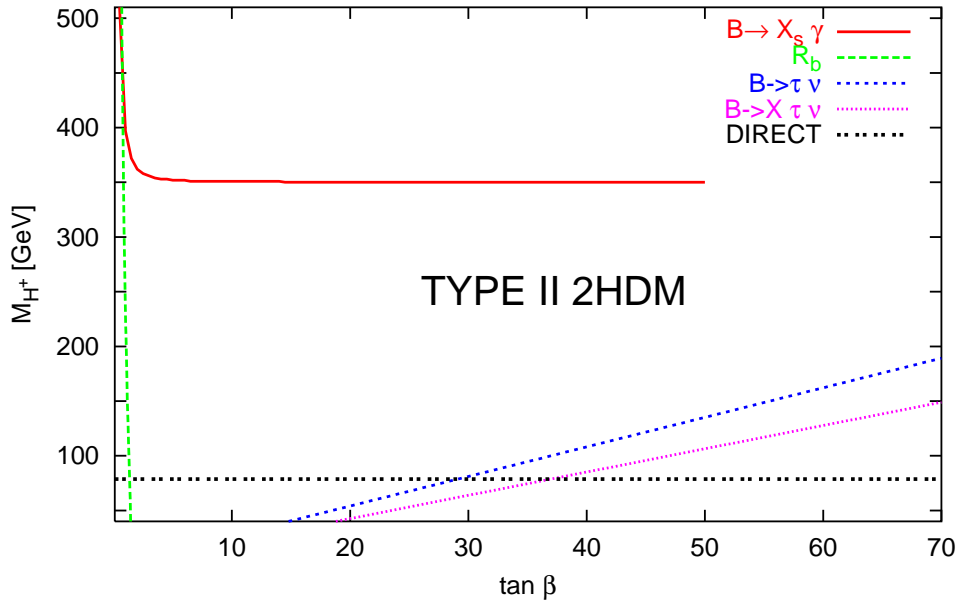}
\caption{
Direct and indirect lower bounds on $M_{H^+}$ from different
processes in type II 2HDM  
as a function of $\tan\beta$~\cite{hurth_rev,gm_0104}.}
\label{2hdm1}
\end{center}
\end{figure}

Fig.~\ref{2hdm1} shows constraints on the $\tan \beta - m_H$ plane
resulting from $B \to X_s \gamma$ along with other
processes. The sensitivity of $B \to X_s \gamma$ is
very impressive. Note that $B \to \tau \nu, X \tau \nu$
cannot compete with $B \to X_s \gamma$ unless $\tan \beta$ is very large;
see\cite{gm_0104} for further details. 

\begin{figure}[h]
\begin{center}
\hskip -1in
\epsfxsize=5in
\epsfbox{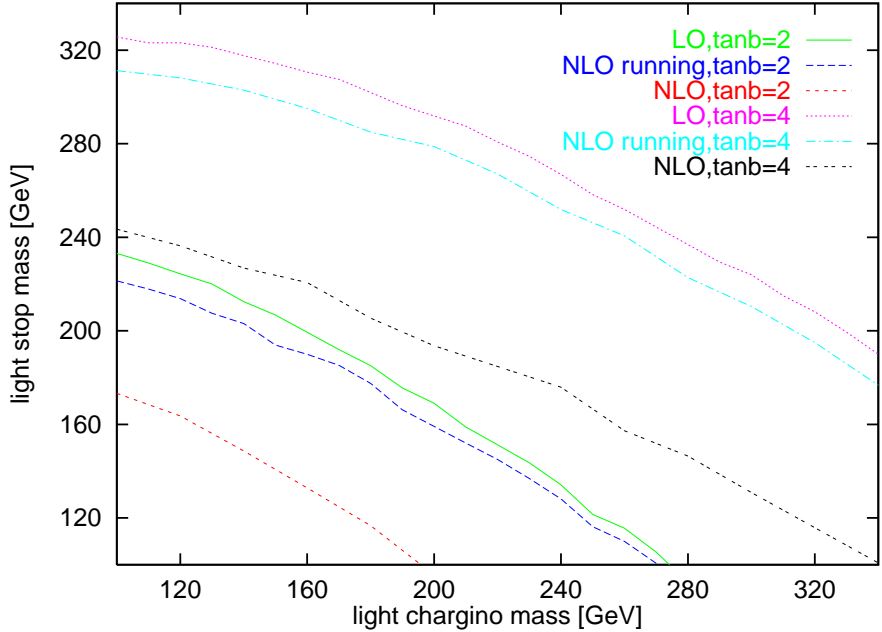}
\caption{
Upper bounds on the lighter chargino and stop masses
from $B \rightarrow X_s \gamma$ data  in a scenario
with a light charged Higgs mass;
for  $\tan\beta=2$ (three lower curves) and $4$
(three upper plots)
the LL, NLL-running and NLL results
(from the top  to the bottom) are shown see~\cite{hurth_rev,mc_98}}. 
\label{susy1}
\end{center}
\end{figure}

The next example Fig.~\ref{susy1} illustrates
constraints on the SUSY parameter space of stop-chargino
masses.

\vskip 0.5in

\begin{figure}[h]
\begin{center}
\hskip -1in
\epsfxsize=5in
\epsfbox{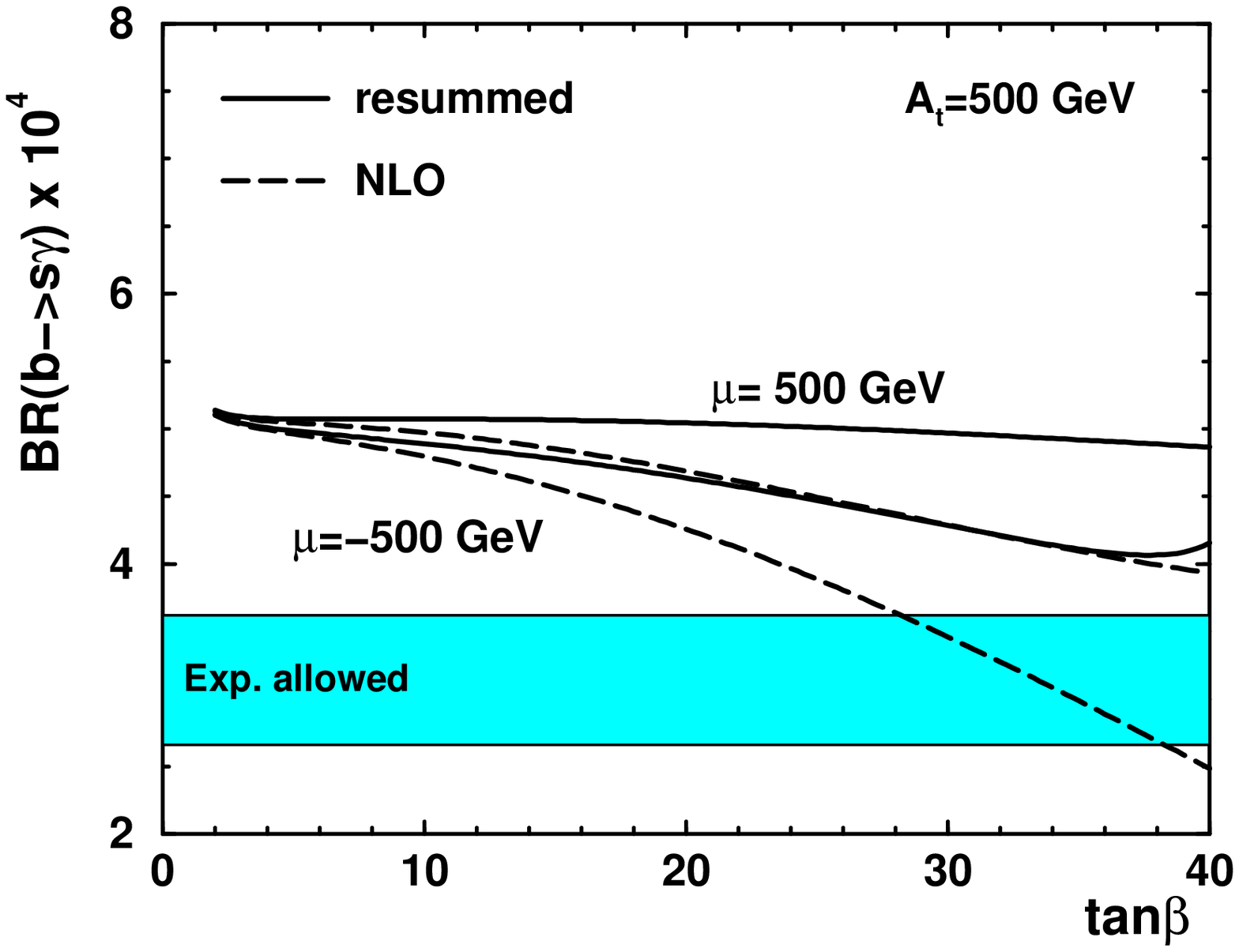}
\caption{
Comparison of the theoretical NLL predictions within
a special MSSM scenario
{\it with} the resummed large $\tan \beta$ terms;
the charged Higgs boson mass is $200$ GeV and the light stop mass is
$250$ GeV.
The values of $\mu$ and $A_t$ are indicated in the plot, while
the gluino, heavy stop and down-squark masses are set at $800$ GeV;
see~\cite{hurth_rev,mc_00}}. 
\end{center}
\end{figure}



\subsubsection{Direct CP violation in Radiative B decays in and beyond
the SM}

As discussed above even if improved measurement of the branching fraction
do not lead to better constraints on the parameter space of
BSM's, direct CP asymmetry can still be a very powerful way
to search for new physics; see Table~\ref{ksw_tab}.
In the $B \to X_s \gamma$ mode,
asymmetries much larger than the SM are possible in
SUSY extensions. However, due to an accidental
cancellation\cite{ksw_0006}
in a class of two Higgs doublet models
the asymmetries do not change much from the SM.

For the more suppressed $b \to d$ transition, the direct CP
asymmetry in the SM is expected to be much larger,
around 16\%. In this mode, new (BSM) CP-odd phase(s)
can decrease or increase the asymmetry compared to 
SM expectations.

Given that the asymmetry in $b \to d$ is
some 15 times bigger than in $b \to s$, even though the
BR of $b \to d$ is expected to be smaller by roughly the same
factor, detection of the $b \to d$ asymmetry may well
require fewer number of B's\cite{ksw_0006}. Recall that a
precise measurement of the branching fraction for the 
mode $B \to \rho (\omega) \gamma$
is also a good way of better determining  
$\frac{|V_{td}|}{|V_{ts}|}$. Eventually, this will require
careful, precise calculation of the SU(3) breaking effects
for the corresponding form factors presumably using lattice methods.

The importance of
studying radiative inclusive and exclusive modes at SBF can
hardly be overemphasized.


\begin{table}
\vspace{0.3in}
\begin{tabular}{|c|c|c|}
\hline
Model & $\acps (\%)$ & $\acpd (\%)$ \\
\hline
SM & 0.6 & -16 \\
\hline
2HDM (Model II) & $\approx 0.6$ & $\approx -16$ \\
\hline
3HDM & -3 to +3 & -20 to +20 \\
\hline
T2HDM & $\approx$ 0 to +0.6 & $\approx$ -16 to +4 \\
\hline
Supergravity\cite{tg_9812}& $\approx$ -10 to +10 & -(5 - 45) and (2 -
21) \\
\hline
SUSY with squark mixing\cite{cc_9808}& $\approx$ -15 to +15 &    \\
\hline
SUSY with R-parity violation\cite{ec_0005}& $\approx$ -17 to +17 &
\\
\hline
\end{tabular}
\caption{Direct CP asymmetries in radiative b decays in and beyond the
SM~\cite{ksw_0006}.}
\label{ksw_tab}
\end{table}

\subsection{B-decays to lepton pairs ($B \to ``X" l^+ l^-$)}

Another very interesting rare B mode, whose importance has
been recognized for a very long time\cite{hsw87}, is
$B \to X l^+ l^-$.  Belle and BaBar have recently started
to see this mode with,
BELLE + BABAR $\Rightarrow Br(B \to X_s l^+ l^-)=(6.2 \pm 1.1
^{+1.6}_{-1.3}) 
\times 10^{-6}$\cite{nakao_lp03}.
SM (NLO) predicts $(4.2\pm 0.7) \times 10^{-6}$\cite{hurth_rev}.
In passing, we also recall the
first SM [LO] prediction\cite{hsw87},
$\approx 6\times 10^{-6}$
for $m_t=175$ GeV. Since this mode is somewhat rarer compared to
$B \to X_s \gamma$ its detection took longer. It ought to
be clear though that it also is nevertheless very important:

\begin{itemize}
\item Inclusive ($X_s$, $X_d$), exclusive (K, $K^*$, $\pi$, $\rho$...)
Br's and CP asymmetries will continue to provide valuable
information on SM parameters and
constraints on BSM physics as better data becomes available from B
and Super-B factories.

\item As an example note the special richness of $K^*$ ($\rho$) final
states that provide numerous ($T_N$ even AND odd) CP-violating
observables\cite
{kruger99}; see Table~\ref{kruger_tab}.

\item While an exhaustive study may well be beyond the reach of even
a Super-B factory,  very clean predictions of the CKM
paradigm\cite{Ali-Hiller98}

$A_{CP}^{X_s} = -(0.19^{+0.17}_{-0.19})\%$;
$A_{CP}^{X_d}=(4.40^{+3.87}_{-4.45})\%$

should certainly be an important target of B-facilities.

Note also the interesting predictions for the corresponding exclusive
modes\cite{kruger99}, $B \to K^* (\rho) l^+ l^-$, given in 
Table~\ref{kruger_tab}.

\end{itemize}    

\begin{table}
\vspace{0.3in}
\begin{tabular}{ccccccccc}
&$A_{CP}$&$A_3$&$A_4$&$A_5$&$A_6$&
$A_
7$&$A_8$&$A_9$\\
\hline
$K^*$&
$2.7$ & $-.6$ & $-2.0$ & $5.2$ &  $-4.6$ & $0$ & $.6$ & $-.04$\\
\hline
$\rho$ &
$-1.7$ & $.1$ & $.4$ & $-1.4$ & $1.2$  & $0$ & $-.1$ & $.006$\\
\hline
\end{tabular}
\caption{Estimates of the average CP-violating asymmetries
$A_k$ in units of $10^{-4}$ ($10^{-2}$) for
the $B\to K^*$ ($B\to \rho$) transition 
\cite{kruger99}. 
Note also that 7-9 are triple
correlation ($T_N$ odd) asymmetries,
others are $T_N$ even; $A_{CP}$ is PRA }. 
\label{kruger_tab}
\end{table}

\subsection{Tree Dominated (hadronic) FS: e.g $\jpsi K (K^*)$}


These decay modes are
extremely sensitive to $\chi_{BSM}$ from an extended scalar
sector\cite{ws99}. In addition, as~\cite{ws99}
emphasizes, they have the significant advantage of
possessing a very clean experimental signal in neutral as well
as in charged B decays and also have large BRs.
CLEO, BELLE and BABAR~\cite{psi_k} have looked for these non-SM
effects. For example, for the $B^{\pm} \to \jpsi K^{\pm}$
mode, the PRA is found to be less than around 5\%.
Clearly it is important to improve these bounds
to look for a BSM-CP-odd phase, especially one 
from charged Higgs exchange. 

To illustrate how such effects
may arise, we may consider a ``Two Higgs doublet
model for the top quark(T2HDM)"\cite{dk96,ksw98}, which is
very well motivated.
\begin{enumerate}
\item{} In this model the large $m_{top}$ value is accommodated naturally by
postulating that the second Higgs doublet,
with a much larger VEV compared to the 1st, couples
only to the top quark giving rise naturally to
$\tan \beta >> 1$.

This is accomplished via the Lagrangian:

\begin{equation}  
L_{yukawa} = -\bar L_L \phi_1 E l_R - \bar Q_L \phi_1 F d_R
-\bar Q_L \bar \phi_1 G {\bf I^{(1)}} u_R -\bar Q_L \bar \phi_2 G
{\bf I^{(2)}} u_R + H.c. \\
\end{equation}   

Here $\phi_{1,2}$ are the two Higgs doublets; E, F and G are 3 x 3
Yukawa matrices giving masses respectively to the charged leptons, the
down and up type quarks;
${\bf I^{(1)}} \equiv diag(1,1,0)$
and ${\bf I^{(2)}} \equiv diag(0,0,1)$
are the two orthogonal projectors onto the 1st two and third family
respectively. $Q_L$ and $L_L$ are the usual left-handed
quark and lepton doublets.

\item{} It is best to view the
T2HDM as a low energy effective theory (LEET) that parameterizes through
the Yukawa interactions some high energy dynamics,
which generate the top quark mass as well as
the weak scale.

\item{} In addition to large $\tan \beta$ the model has
restrictive FCNC (since it belongs to type III) amongst
only the up-type quarks.

\item{} A distinctive feature is also
that $b \to c$ couplings becomes complex with non-standard
CP violation in many B-decays, including the ``gold-plated"
mode, $B^0 \to \jpsi K^0$. B-factory measurements now
imply that such a non-standard phase is subdominant.

\item{} A good way to search for the presence of a
small $\chi_{BSM}$ is to search for direct CP-asymmetry
in the experimentally clean channel, $B^{\pm} \to
\jpsi K^{\pm}$ where the SM predicts completely
negligible PRA\cite{ws99}.

\item{} A possible drawback of PRA in $\jpsi K$ is
that the needed strong-phase {\it may} also not be 
large. For this reason it is very important also to study TCA
in $B \to \jpsi K^*$\cite{lss0207}

\item{} It is important to understand that the presence
of complex (non-standard) (tree) couplings (e.g. $b \to c$)
also has important consequences for $b \to s(d)$ penguins.
Thus penguin dominated hadronic FS as well as radiative,  
pair leptonic and semi-leptonic FS become useful testing grounds.
\end{enumerate}

\subsection{Constraining the Higgs sector with $B \to D(^*) \tau
\nu_{\tau}$}

Semi-leptonic B decays involving $\tau$ leptons
provide important avenues to search for BSM physics,
especially of the type involving an 
extended charged Higgs sector, CP-conserving or CP violating. 
The T2HDM discussed in the preceding section
is a fine example of such a model. Below are the important points,

\begin{enumerate}
\item{} The $q^2$ ($q \equiv p_B-p_D$) distribution
is rather sensitive to $\frac{\tan \beta}{m_H}$, much more so than
the integrated rate.
\item{} Unlike $B \to X_s \gamma$, in $B \to D(^*) \tau \nu_{\tau}$
the charged Higgs contribution does not cancel against
the contribution from other SUSY partners.
\item{} The transverse polarization
$(p_{\tau}^t)$ of the $\tau$ is an extremely sensitive
and uniquely clean probe of a CP-odd $H^\pm$ phase;
$<p_{\tau}^t>_{SM} = 0$.
\end{enumerate}

\subsubsection{(CP-conserving) Constraints on Higgs sector with $B \to D
\tau \nu_{\tau}$}

This reaction has received considerable attention in the past several
years\cite{hs91,t95,ghn95,ks97}. Note that:

\begin{enumerate}
\item{} The needed semi-leptonic form factors (3 of them: $F_0,F_1,F_s$)
should be determined accurately via $B \to D e \nu$ and
$B \to D \mu \nu$ \\

\item{} For illustrative purposes we invoke Heavy Quark Symmetry (HQS), 
which gives all 3 form factors in terms of the Isgur-Wise
function.\\

\item{} Although $\frac{1}{m_Q}$ corrections to individual form factors
are appreciable, their ratios receive very small residual corrections.
We use \cite{mn_pr} and parameterize the remaining errors \\
\end{enumerate}

\begin{figure}[h]
\begin{center}
\epsfxsize=5in
\hskip -1in
\epsfbox{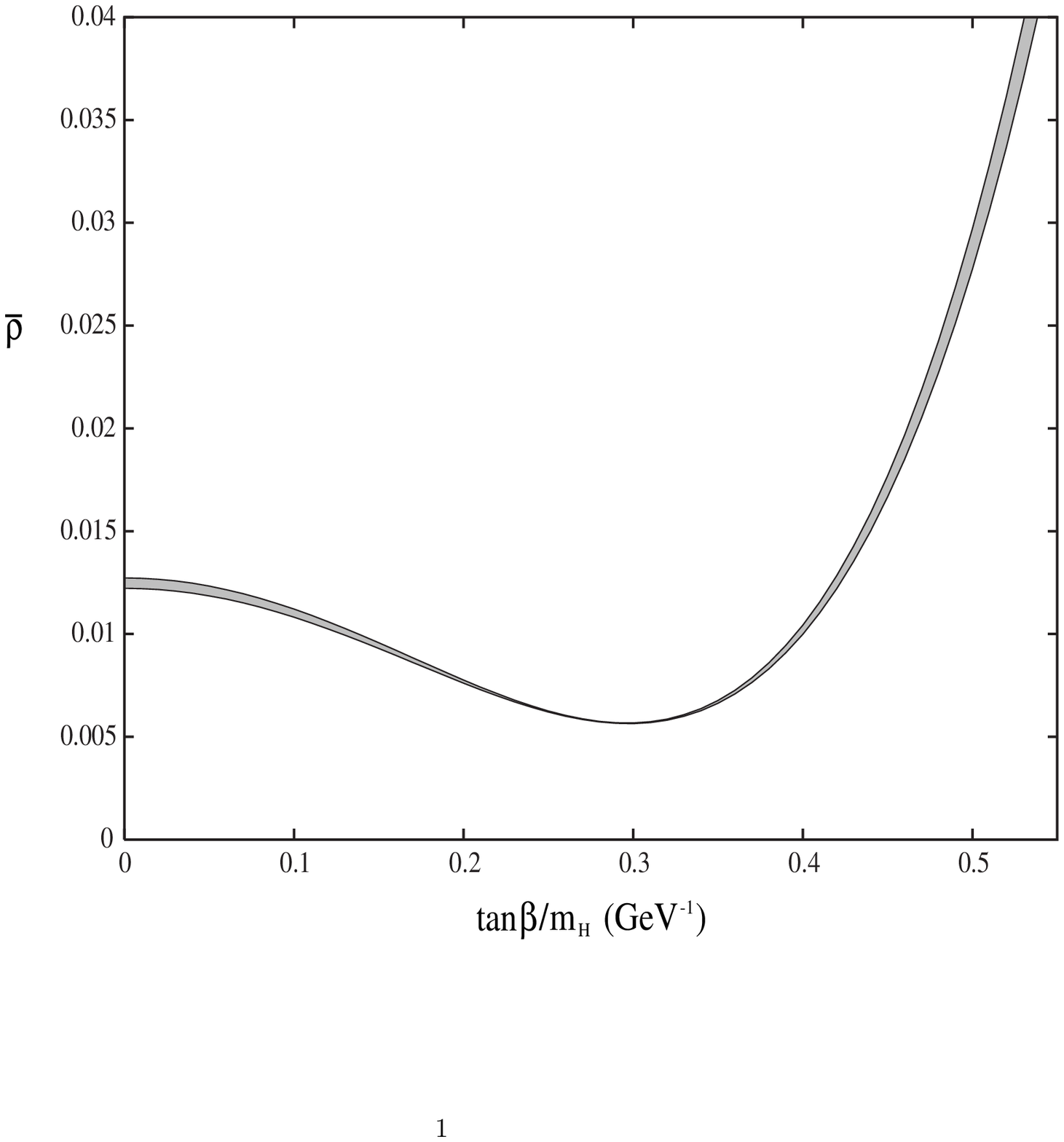}
\caption{Normalized integrated width for $B \to D \tau \nu$ 
as a function of $\tan
{\beta}/m_H$. The shaded region again indicates
the theoretical uncertainty originating from the 
uncertainty in the form-factors\cite{ks97}.  }
\label{diff_sl}
\end{center}
\end{figure}

Theoretical study~\cite{ks97} shows that the 
differential spectrum is clearly more
sensitive (see Fig.~\ref{diff_sl}).
Note [GeV $tan{\beta}/m_H$] $= [0,0.1,0.3]$ for $\bar {\rho}/\bar {\rho}_{SM}
=1, 80\%, 50\%$,  respectively. The 
differential rate is given by, \\

\begin{equation}   
\frac{d\Gamma_{\tau}}{dt} = f(r_{\tau}, r_D, t, \zeta(t), \delta_H(t)),
\end{equation}  

where, $r_D = \frac{m_D^2}{m_B^2}$, $r_\tau = \frac{m_{\tau}^2}{m_B^2}$,
$t = \frac{q^2}{m_B^2}$

\begin{equation}  
\delta_H(t) = -[\frac{\tan \beta}{m_H}]^2 \frac{m_bm_B^2t}{m_B-m_D}
[1 + \frac{m_c}{m_b} \cot ^2 \beta] \frac {F_s(t)}{F_0(t)}
\end{equation}  


\begin{figure}[h]
\begin{center}
\epsfxsize=5in
\hskip -1in
\epsfbox{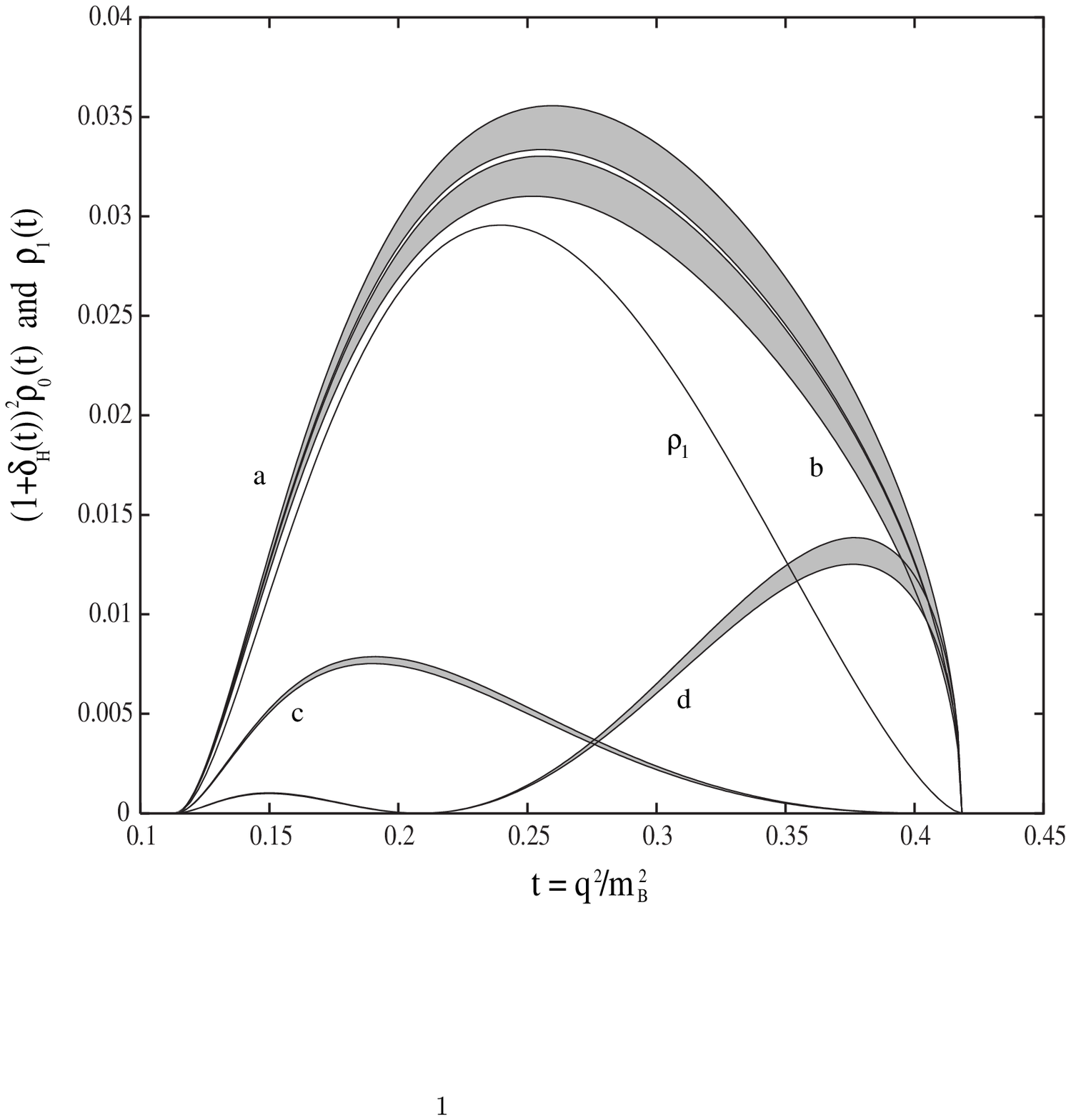}
\caption{Spin-0 and spin-1 contributions to the
differential distribution for $B \to D \tau \nu$
in the Standard Model and in two-Higgs
doublet models with 
[GeV $\frac{tan \beta}{m_H}$] $=$ 0 (the SM); 
0.06; 0.25; 0.35 for curves a; b; c; d
respectively. The shaded regions indicate the theoretical uncertainty
due to the uncertainties in the form-factors. The solid curve
corresponds to the spin-1 contribution, $\rho_1(t)$ 
~\cite{ks97}.}
\label{higgs_sl}  
\end{center}
\end{figure}

It is difficult to go below 0.06 in t (see Fig.~\ref{higgs_sl})
due to residual uncertainties in the ratio of form factors.


One can also try to improve the 
constraints by use of the technique of
optimized observables~\cite{as92,ggh96}.
Effects of QCD corrections have also been studied~\cite{ghn95}.


\subsubsection{Transverse $\tau$ polarization in $B \to \tau \nu_\tau
X$}

This is an extremely sensitive observable for
probing the presence of a CP-odd phase ($\chi^{H^\pm}_{BSM}$)
from charged Higgs exchange present in {\it e.g.} the T2HDM
discussed in the preceding sections.   
\noindent Recall also that
due to CPT, CP violating observables can be split into 2
categories\cite{abes_pr}.
\begin{itemize}
\item{} $T_N$ even, (e.g. $<E_\tau>$ or PRA) $\Rightarrow \propto$ Im
Feynman amplitude 
i.e. $\sin{\delta_{st}}$; $\delta_{st}$ is the CP-even "strong" phase.
\item{} $T_N$ odd , (e.g.$<p_{\tau}^t>$) $\Rightarrow \propto$ 
Re Feynman  
amplitude i.e. $\cos{\delta_{st}}$\\
\end{itemize}
where,

\begin{equation}  
p_{\tau}^t \equiv \frac{S_\tau \cdot p_\tau \times p_X}{|p_\tau \times
p_X|}.  
\end{equation}  

Thus,
$<E_\tau>, A_{PRA}$ require an imaginary part of a Feynman amplitude
and are proportional to
$\frac{\alpha_s}{\pi} \approx 0.1 $
Also, for $<E_\tau>, A_{PRA}$, W-H interference requires
the amplitude to be proportional to
$Tr[\gamma_{\mu}L({\slp}_{\tau}+m_{\tau})(L,R){\slp}_{\nu}]$.
This yields another suppression factor,  $\propto m_{\tau}/m_B$.
Therefore, $\frac {<p^t_\tau>}{<E_\tau>}$ or
$\frac {<p^t_\tau>}{A_{PRA}} \approx 30$
\cite{aes93}.
The effect of
power corrections was studied in \cite{gl94} and tends
to somewhat reduce this enhancement.  


\vskip 0.5in
Experimental detection of $P_{\tau}^t$, via decay
correlation in $\tau \to \pi \nu, \mu \nu \nu, \rho \nu$ etc.
is expected to be much harder than measuring an  energy or rate asymmetry.
Clearly rate and/or energy asymmetries should also be studied
especially if detection efficiencies for those are higher.
Fake asymmetries due to FSI can arise if only $\tau^-$ or
$\tau^+$ is studied. Genuine (i.e. CP violating) 
$p_{\tau}^t$ will switch sign from $\tau^-$ to $\tau^+$. 
%

Although from a theoretical standpoint, these semi-leptonic modes
with $\tau$ in the final states are rather unique and extremely
clean, their experimental study is a very difficult challenge.
The main problem is that due to large backgrounds, at the moment, 
the only way to see these modes is with the use of fully reconstructed
tagged events. Unfortunately the tagging efficiency is
only ${\cal O}(0.4)~\%$\cite{belle_loi}. 
Combining this with the detection efficiency 
and the branching ratio ends up leaving too few events
to have a serious impact on the allowed parameter space.

\begin{table}
\begin{tabular}{|c|c|c|c|}
\hline
Final State & Observable & Theoretical Cleanliness & Sensitivity to NP\\
\hline
$\gamma [K_s^*,\rho,\omega]$ & TDCP & 5* & 5*  \\
\hline
$K_s[\phi,\pi^0,\omega,\eta',\eta,\rho^0]$ & TDCP & 4.5* &  5* \\
$K^*[\phi,\rho,\omega]$ & TCA &  4.5* &  5* \\
\hline
$[\gamma,l^+l^-] [X_s,X_d]$ & DIRCP & 4.5* & 5*  \\
\hline
same   & Rates & 3.5* & 5* \\
\hline
$\jpsi K $ & TDCP, DIRCP & 4* & 4*  \\
\hline
$\jpsi K^* $ & TCA &  5* & 4*  \\
\hline
$D (*) \tau \nu_{\tau}$ & TCA ($p_t^{\tau})$ & 5* & 4*  \\
same & Rate & 4* & 4* \\
\hline
\hline
\end{tabular}
\caption{Final states and observables in B - decays
useful in searching for effects of New Physics. Reliability of
SM predictions and sensitivity to extensions
of the SM are each indicated by stars ($5 = best$)}. 
\label{dir_seek_tab}
\end{table}

\section{Crucial benchmarks in the hunt for $\chi_{BSM}$}

In the hunt for $\chi_{BSM}$ and NP a very good strategy may be to
aim for some specific targets.
Below are a representative sample:

\begin{itemize}
\item Determination of all three angles of UT with errors $ \approx
O(ITE)$
\item Precise determination of $a_{CP} (B \to X_s \gamma)$
(the SM expectation is around 0.6 \%)\\
\item Precise determination of $a_{CP} (B \to X_s l^+ l^-)$
(the SM expectation is $< 0.5\%$)\\
\item Precise determination of Br ($B \to X_d \gamma$)
(the SM expectation is around $10^{-5}$)\\
\item Precise determination of $\sin 2 \beta$, in
penguin dominated final states, {\it i.e.}
$(\phi, \eta' , \pi^0 , \rho^0, \omega) K_s$\\ 
\item Precise determination of TDCPA (S) for $K_s^* \gamma$
(the SM expectation is $\approx 3 \%$)\\
\end{itemize}
The Super-B Factory should be able to meet many if not all of these
goals. Through such a strategy, a SBF would provide several approaches  
to uncovering $\chi_{BSM}$, irespective of
what the underlying theory is.

\section{Summary and Outlook}
B-factories have started on an important hunt.  
One crucial milestone has already been
attained. Not only is the KM phase confirmed,
its dominant role in $B \to \jpsi K^0$ is established !
However, very good theoretical arguments still suggest
that a BSM phase ($\chi_{BSM}$) should exist.
In light of B-factory results its likely that the effects of $\chi_{BSM}$
on B-physics are subtle. Therefore, we will need large numbers of
B's to find $\chi_{BSM}$.
A Super-B Factory with $10^{10}$ B's will allow~\cite{as_izu}:\\
\begin{enumerate}
\item{} Very clean determination of all 3 angles of the UT with errors
around ITE i.e. $O(1\%)$ compared to the current level of (at least) around
$20\%$. This is the most compelling rationale for a SBF
as it will allow a thorough understanding of the CKM paradigm
and the workings of the SM in the flavor sector. Of course,
it is also an excellent way to search for $\chi_{BSM}$.
\item{} Search of small deviations using input from theory
will require a systematic collective effort. Theory (especially lattice)
needs improvement but also continuum methods. If deviations
from the UT due to $\chi_{BSM}$ are not too small,
and are around say $5-10 \%$, then this strategy has a chance.

\item{} More importantly SBF will allow
numerous direct searches for $\chi_{BSM}$ via
\begin{itemize}
\item TDCPA in $\phi K_s$, $\eta' K_s$,
$\pi^0 K_s$, $\rho K_s$, $\omega K_s$, $K^*\gamma$, $\rho \gamma$ .....
\item DIRCPA including TCA in $\phi K^{\pm}$, $\phi K^*$,
$\eta'(\eta) X_{s,d}(K, \rho)$,
$\jpsi K (*)$,\\
$\gamma X_{s,d}(K^*,\rho...)$, $l^+l^- X_{s,d}(K^*,\rho...)$,
$\tau \nu_{\tau} X_c[D(^*)]$
\end{itemize}

\end{enumerate}
Table~\ref{dir_seek_tab} presents a list of the many interesting and
powerful ways to directly search for $\chi_{BSM}$ and NP.
We also indicate there how reliably the SM predictions can be calculated
(indicated by stars with 5 representing the best) and also 
sensitivity to NP, which is again indicated similarly by stars.
The SM predicts negligible asymmetries in many cases of interest.

A Super-B factory will allow constraints 
on $\chi_{BSM}$ to improve by 1-2 orders of
magnitude thereby refining our understanding of flavor-physics
to an unprecedented level. 
In the hunt for $\chi_{BSM}$ several of these provide compelling
benchmarks. Without reaching these our understanding of SM-CKM paradigm
is seriously incomplete. A SBF would
provide multiple possible paths to $\chi_{BSM}$, irrespective of
what the underlying theory is, whether it is SUSY, extra dimensions or 
some altogether different possibility. 
The most tantalizing prize of a Super-B factory is 
the discovery of a  $\chi_{BSM}$, which could 
significantly illuminate our understanding of baryogenesis.

\vskip 1in 
{\bf Acknowledgements}

AS would like to thank the organizers of WHEPP-8,
and especially Uma Sankar, for the 
kind invitation. The work of AS is supported in part by US DOE contract  
No. DE-AC02-98CH10886. The work of TEB is supported in part by 
DOE contract No. DE-FG02-04ER41291.


\end{document}